\documentclass[a4paper,fleqn,usenatbib]{mnras}

\usepackage[T1]{fontenc}
\usepackage{ae,aecompl}
\usepackage{lineno}
\usepackage{threeparttable}

\usepackage{graphicx}	
\usepackage{amsmath}	
\usepackage{amssymb}	

\usepackage{array,etoolbox}
\preto\tabular{\setcounter{magicrownumbers}{0}}
\newcounter{magicrownumbers}

\title[Nature of helicity injection in non-erupting ARs]{Nature of helicity injection in non-erupting solar active regions}

\author[P.~Vemareddy]{
P.~Vemareddy,$^{1}$\thanks{E-mail: vemareddy@iiap.res.in}
\\
$^{1}$Indian Institute of Astrophysics, Sarjapur road, II Block, Koramangala, Bengaluru-560 034, India
}
\date{ Accepted: 2022 August 03; Revised: 2022 July 28; Received: 2022 May 23}

\pubyear{2022}

\begin{document}
\label{firstpage}
\pagerange{\pageref{firstpage}--\pageref{lastpage}}
\maketitle

\begin{abstract}
Using time-sequence vector magnetic field and coronal observations from \textit{Solar Dynamics Observatory}, we report the observations of the magnetic field evolution and coronal activity in four emerging active regions (ARs). The ARs emerge with leading polarity being the same as for the majority of ARs in a hemisphere of solar cycle 24. After emergence, the magnetic polarities separate each other without building a sheared polarity inversion line. In all four ARs, the magnetic fields are driven by foot point motions such that the sign of the helicity injection ($dH/dt$) in the first half of the evolution is changed to the opposite sign in the later part of the observation time. This successive injection of opposite helicity is also consistent with the sign of mean force-free twist parameter ($\alpha_{av}$).  Further, the EUV light curves of the ARs in 94\AA~and GOES X-ray flux reveal flaring activity below C-class magnitude. Importantly, the white-light coronagraph images in conjunction with the AR images in AIA 94 \AA~delineate the absence of associated CMEs with the studied ARs. These observations imply that the ARs with successive injection of opposite sign magnetic helicity are not favorable to twisted flux rope formation with excess coronal helicity, and therefore are unable to launch CMEs, according to recent reports. This study provides the characteristics of helicity flux evolution in the ARs referring to the conservative property of magnetic helicity and more such studies would help to quantify the eruptive capability of a given AR.
\end{abstract}



\section{Introduction}
\label{Intro}
Magnetic helicity (MH) is a physical parameter to measure the structural properties of a magnetic field.  Using this parameter, one quantifies twist, shear, linking, and kinking of the magnetic field \citep{Mofatt1978_book}. MH is conserved to an excellent approximation in the highly conducting plasmas present in the Sun and heliosphere. In addition, the MH can be transported between regions and between different length scales. Especially, helicity transfer between regions involves either the bulk transfer of a helicity-carrying field, or propagation of twist and braiding along field lines crossing the boundary between the regions \citep{Berger1984,pevtsov2014}.

On the Sun, magnetic flux emergence occurs in the localized regions known as active regions (ARs), and are the potential locations of the MH transfer from the convection zone to the atmosphere. In these regions, the evolution of the magnetic field is driven by slow plasma motions. These motions are dominated by vertical components in the early emergence phase and later dominated by horizontal components of velocity (e.g., \citealt{MagaraLongcope2003}). The equation for the helicity transfer through the surface into the open volume like solar corona was derived by \citep{Berger1984_Hcons} and contains shear and emergence terms. Using continuous observations of the magnetic field from Michelson Doppler Imager (MDI; \citealt{Scherrer1995_MDI}), \citet{Chae2001_ObsDetMagHel}, for the first time, estimated the helicity flux transfer rate in the AR 8011. Later several authors followed a similar method to estimate the helicity transfer rate during the AR emergence and later phases of evolution (e.g.,  \citealt{Demoulin2002_MagHelShMot, pariat2005, Labonte2007_SurveyMagHel, tian2008a, vemareddy2012_hinj, Liuyang2012_HelEner, Vemareddy2015_HelEne}) to infer the coronal helicity budget and then relate to the flaring activity. However, the flaring activity is primarily related to the magnetic reconnection, which is sometimes associated with CME, so the occurrence of a CME is of more fundamental and major scientific interest in the space-weather point view. Since the CME is the manifestation of twisted flux rope (FR) eruption \citep{rust1996, Vourlidas2013}, formation and existence of a twisted flux rope in the AR magnetic structure is the key observational aspect to link the photospheric helicity flux transfer. 

\begin{figure*}
	\centering
	\includegraphics[width=.99\textwidth,clip=]{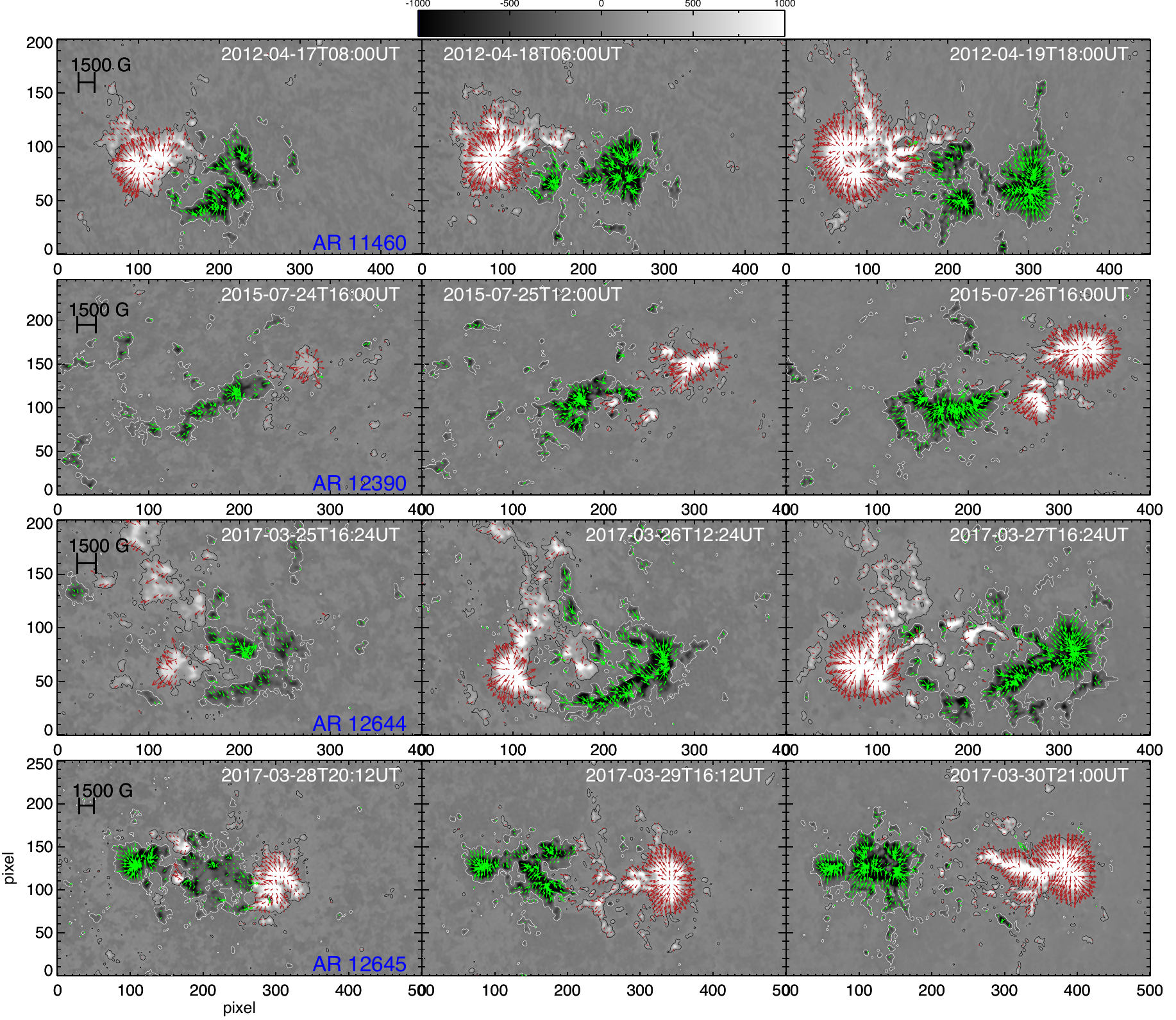}
	\caption{HMI vector magnetic field observations of four ARs (top to bottom) at three different times (left to right) of their evolution. The background image is the normal component of the magnetic field  with contours at $\pm$120 G. Arrows refer to horizontal fields, ($B_x$, $B_y$), {\bf which are scaled to 1500 G as shown in left column panels}. The axes units are in pixels of 0.5 arcsec.  }
	\label{Fig1}
\end{figure*}

\begin{table*}
	\caption{Details of the Active Regions in this study}
	\begin{threeparttable}
	\centering
	\begin{tabular}{lllllll}
	\hline \hline
	s. no & NOAA &  Latitude\tnote{a}  & Duration\tnote{b} & sign ($dH/dt$)\tnote{c} & activity\tnote{d}        \\
	&   AR    &     &     &     &     \\
	\hline\hline 

1 & 11460 & N16  & 17--22 Apr, 2012  &  +/-ve	 &  4 C-flares \& no CMEs   \\
2 & 12390 & S15  & 24--29 Jul, 2015  &  +/-ve   &  B-flares \& no CMEs   \\ 
3 & 12644 & N14 & 25--30 Mar, 2017  &  +/-ve   &  2 C-flares \& no CMEs    \\
4 & 12645 & S09  & 29 Mar--03 Apr, 2017 &  -/+ve & 10 C-flares \& No CMEs     \\
\hline
\end{tabular}
\begin{tablenotes}
\item[a] Latitudinal position of the AR; north (N)  or south (S)
\item[b] period of the AR in this study during its disk transit within $\pm$40$^\circ$ longitude
\item[c] sign of helicity flux over time
\item[d] observed GOES X-ray flares ($>$C1), LASCO-C2 CMEs
\end{tablenotes}
\label{tab1}
\end{threeparttable}
\end{table*}

From the conservation property of the helicity, \citet{zhangmei2005,Zhangmei2008_HelBound} proposed that an already formed twisted flux rope (twisted flux) with continuous injection of MH inevitably erupts in order to remove the excess accumulated coronal helicity. First, twisted FR forms by shear and rotating motion of magnetic foot points which implies continuous injection of magnetic helicity of a predominant sign. Second, the eruption requires further injection of helicity upto a point of upper bound that the FR can remain in equilibrium. Studies of long-term AR evolution showed that the magnetic field evolution with continuous injection of helicity with a predominant sign exhibits presence of twisted FR as the sigmoidal structures and their eventual eruption as CMEs \citep{liuchang2007,jiangc2014, Vemareddy2016_sunspot_rot,Vemareddy2017_homologus,Vemareddy2018a}. However, the estimation of the upper bound for a given magnetic flux distribution for the prediction of a CME occurrence is still a formidable task, which leads to questionable Low's conjuncture of helicity upper bound.  

\begin{figure*}
	\centering
	\includegraphics[width=.9\textwidth,clip=]{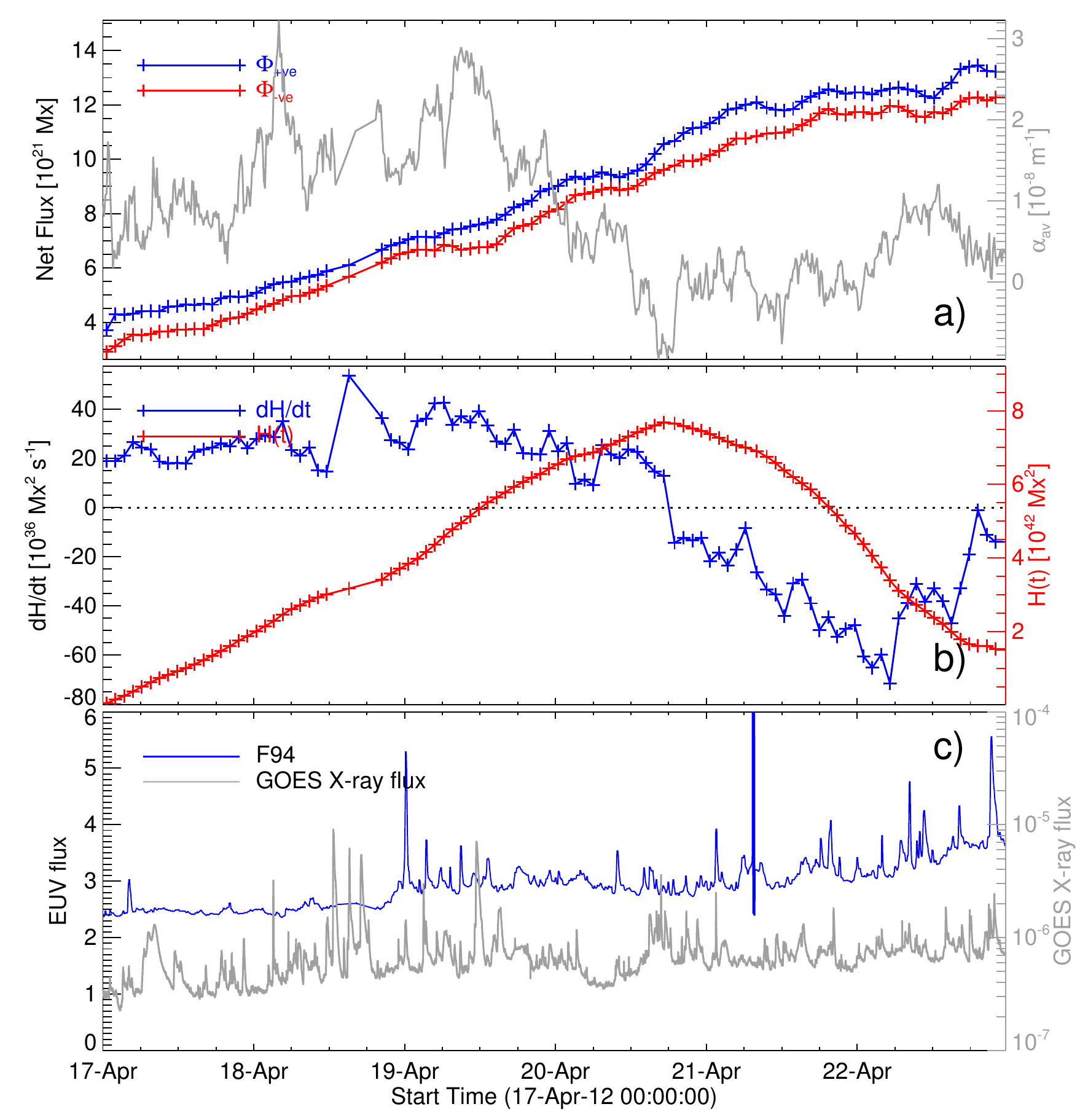}
	\caption{Time evolution of AR 11460, a) Net magnetic flux showing increasing flux content as the AR emerges and evolves in time with separating magnetic polarities. $\alpha_{av}$ (grey) is also plotted with y-axis scale on the right. b) the helicity injection rate ($dH/dt$). Time accumulated quantity $H(t)$ (red) is also plotted with Y-axis scale on right side, and c) Light curves of the AR in EUV intensity 94~\AA~waveband. GOES flux is also shown with y-axis scale on right.  }
	\label{Fig_evol_11460}
\end{figure*}

Alternatively, in order to support the Low's conjecture, the recent studies search for the ARs with helicity injection changing sign over time. In such cases, the coronal helicity accumulation in the first part of the evolution is possibly canceled by the later part. Therefore, the AR magnetic structure can not reach the state of the helicity upper bound and is incapable of launching CMEs. Detailed study of the emerging ARs 11928, 12257, the helicity flux was found to change sign from positive to negative over its evolution \citep{Vemareddy2017_OppoHel, Vemareddy2021_OppoHel}. Importantly, CMEs were not launched from these ARs except the C-class flaring activity. Also, a twisted flux rope structure was not present as inferred by the non-linear force free magnetic field extrapolations. Such examples of the ARs provide evidence for the important role of conservation property of the MH in generating the eruptive activity. In this paper, we present observations of four ARs exhibiting changing sign of helicity flux transfer over time. In Section~\ref{Obs}, observational data and analysis procedures are described, results are presented in Section~\ref{Res} and conclusions are given in Section~\ref{Dis}.

\begin{figure*}
	\centering
	\includegraphics[width=.9\textwidth,clip=]{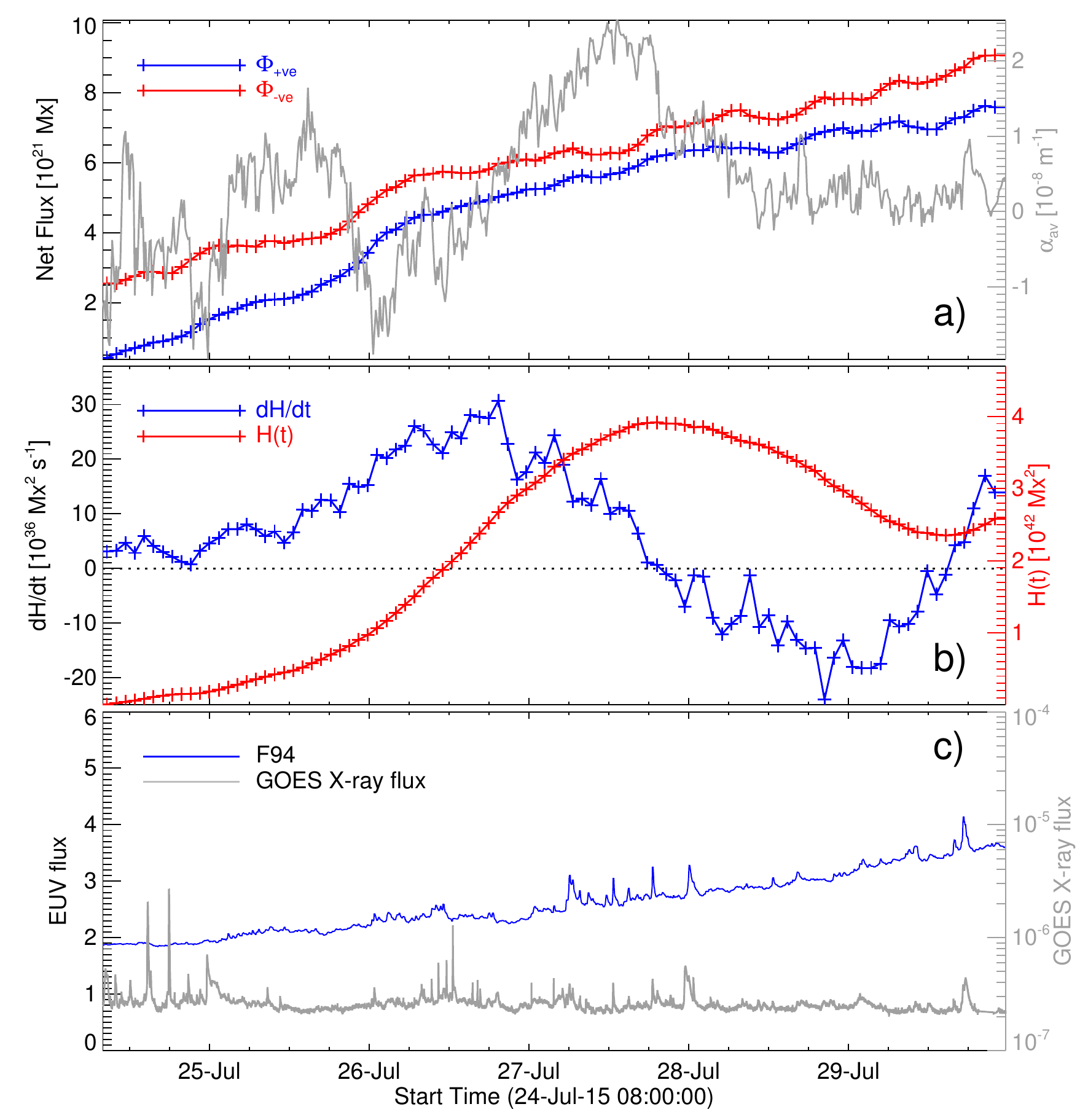}
	\caption{Same as Figure~\ref{Fig_evol_11460} but for AR 12390.}
	\label{Fig_evol_12390}
\end{figure*}

\begin{figure*}
	\centering
	\includegraphics[width=.9\textwidth,clip=]{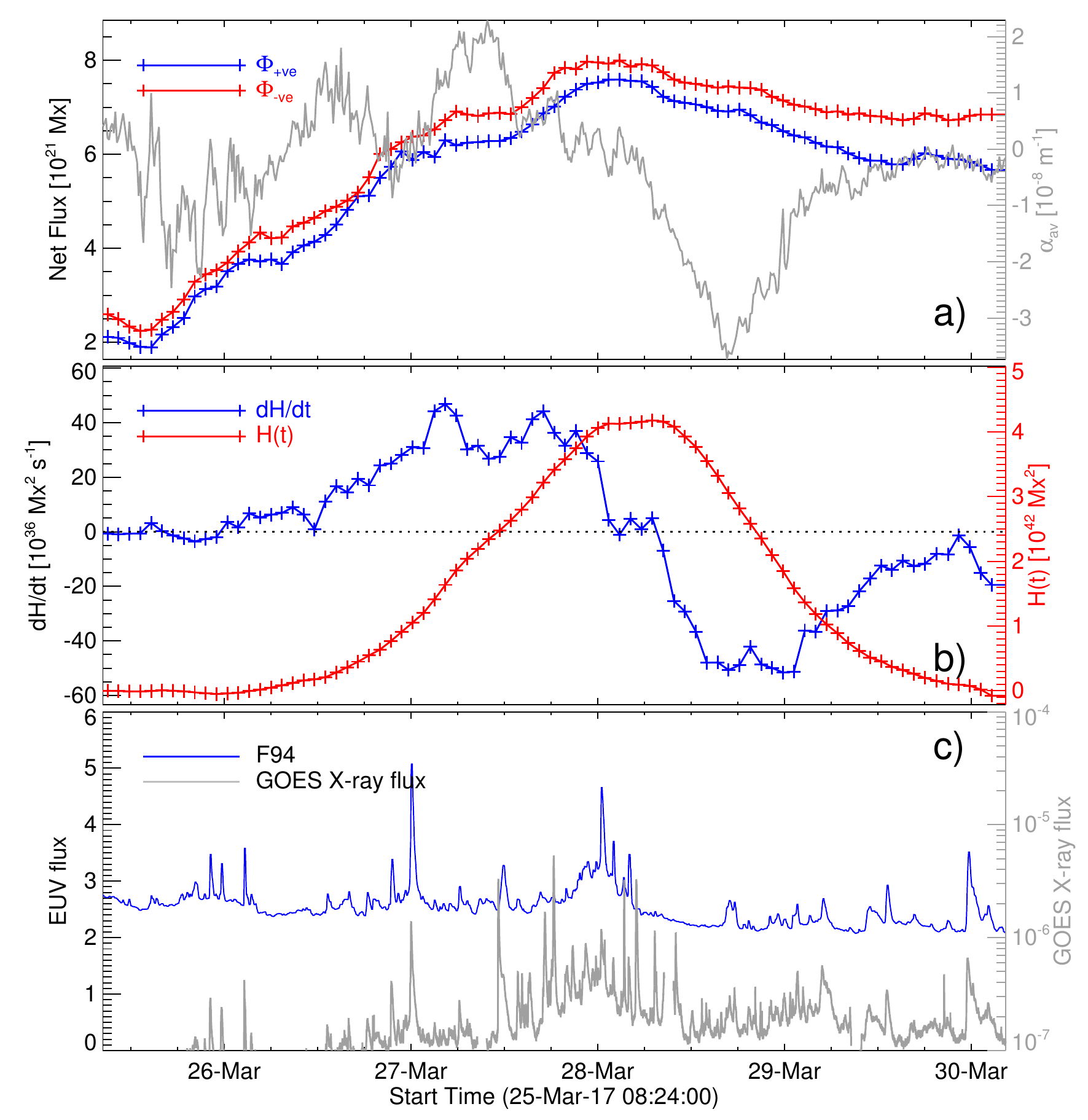}
	\caption{Same as Figure~\ref{Fig_evol_11460} but for AR 12644.} 
	\label{Fig_evol_12644}
\end{figure*}

\section{Data and analysis procedures}
\label{Obs}
In this study of helicity flux evolution in the ARs, we used uninterrupted high-resolution photospheric vector magnetic field observations obtained by the \textit{Helioseismic and Magnetic Imager} (HMI, \citealt{schou2012}) on board Solar Dynamics Observatory. The HMI observes the full solar disk in the Fe {\sc i} 6173\AA~spectral line at a spatial resolution of 0''.5/pixel. About the line center, full disk filtergrams are obtained at six wavelength positions to compute Stokes parameters I, Q, U, and V which are then reduced with HMI science data processing pipeline \citep{hoeksema2014} to retrieve the vector magnetic field. We used data product (\texttt{hmi.sharp.cea\_720s}) at 12 minute cadence containing the field components $(B_r, B_\theta, B_\phi)$ in heliocentric spherical coordinates which are approximately equal to $(B_z, -B_y, B_x)$ in Heliographic Cartesian coordinates. To limit the involved project effects, the emerging ARs of interest are chosen within $\pm40^\circ$ longitude. In Figure~\ref{Fig1}, we displayed the example maps of the vector magnetogram observations of the four ARs from early to post emergence phase. In this figure, the horizontal magnetic field (arrows) is plotted on the map of the vertical component of the magnetic field. As the ARs evolve in time, the opposite magnetic polarities grow in size with their separation being increased. 

From the these vector magnetogram observations of the ARs, the velocity field $\mathbf{v}$ is derived from DAVE4VM procedure \citep{schuck2008}. We then computed the helicity injection rate \citep{Berger1984_Hcons} by

  \begin{equation}
  {\left. \frac{dH}{dt} \right|}_{S} =
     2\int_{S}{\left( {{\mathbf{A}}_{p}}\centerdot {{\mathbf{B}}_{t}} \right)  {{\text{v}}_{n}} \,\rm d\mathbf{S}}
    -2 \int_{S}{\left( {{\mathbf{A}}_{p}}\centerdot {{\mathbf{v}}_{t}} \right){{B}_{n}} \,\rm d\mathbf{S}}
  \label{EqHel}
  \end{equation}
where subscript $t$ represents the transverse components and $n$ represents the normal component of $\mathbf{v}$ and $\mathbf{B}$, respectively. $\mathbf{A}_p$ is the vector potential of the coronal potential magnetic field which is computed with $B_n$ as the photospheric boundary condition by imposing the Coulomb gauge condition, $\nabla \cdot \mathbf{A}_p=0$.

To compare the sign of the $dH/dt$, we also compute the mean force-free twist parameter \citep{hagino2004}

\begin{equation}
\alpha_{av} = \frac{\int J_n \, {\rm sign}(B_n) \,\rm d\mathbf{S} } { \int |B_n| \,\rm d\mathbf{S}}
\end{equation} 
which is also a measure of the non-potentiality of the AR magnetic field. Additionally, we derived the light curves of the EUV emission of the AR in 94 \AA~obtained from the \textit{Atmospheric Imaging Assembly} (AIA, \citealt{Lemen2012}). For this, the AR patches are tracked in time at a time cadence of 3 minutes for the entire duration of the study. 

\begin{figure*}
	\centering
	\includegraphics[width=.9\textwidth,clip=]{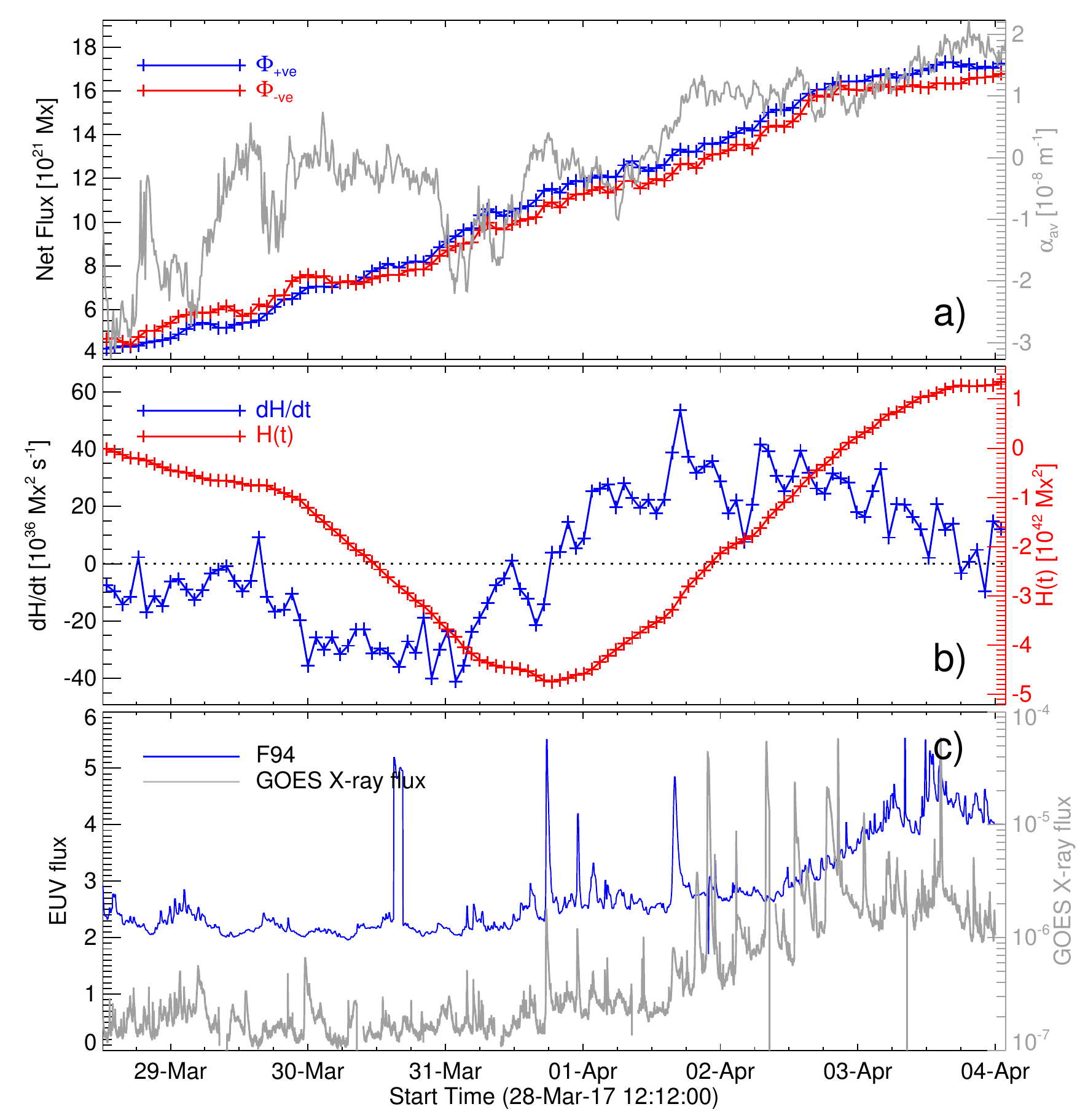}
	\caption{Same as Figure~\ref{Fig_evol_11460} but for AR 12645.} 
	\label{Fig_evol_12645}
\end{figure*}

\section{Results}
\label{Res}
To ascertain the role of helicity flux in the generation of eruptive events, we need to know the coronal helicity budget which is not known for already emerged AR. Therefore, we target the emerging ARs within the visible disk. We found very few cases of ARs exhibiting sign change behavior of injected helicity while the AR emerges and evolves. In the following, we discuss the evolution of $dH/dt$ and the accompanying activity in four ARs as listed in the Table~\ref{tab1}. 

\subsection{AR 11460}
The AR 11460 started emerging on April 16, 2012 at N15$^o$E26$^o$. We consider the AR observations between April 17 to 22  The leading polarity is negative consistent with the most ARs in the northern hemisphere of solar cycle 24. The time evolution of various parameters in this AR is shown in Figure~\ref{Fig_evol_11460}. The evolution of $dH/dt$ is positive until the end of April 20, which then turns negative thereafter. This sign change of $dH/dt$ is also consistent with the average force-free twist parameter $\alpha_{av}$ as shown in Figure~\ref{Fig_evol_11460}a. The accumulated helicity budget in corona will be monotonously increasing until the sign of $dH/dt$ is positive, which then decreases when the sign of $dH/dt$ becomes negative. This kind of $dH/dt$ evolution is a signature that the magnetic structure has not reached the state of critical non-potential state. The light curve of EUV emission of the AR in AIA 94 is plotted in Figure~\ref{Fig_evol_11460}c along with the GOES soft X-ray flux. The peaks in the AIA 94 light curve represent the flaring activity in the AR some of which are co-temporal with the GOES X-ray flux. It shows that the AR produced only C-class flares not associated with CMEs.  

\subsection{AR 12390}
Another AR 12390 in our study emerged at E30$^o$S15$^o$ on July 24, 2015. Following the Hale polarity law, the leading polarity of this AR is positive similar to other ARs in the southern hemisphere. The flux content in the AR grows to a moderate level of order $10^{22}$ Mx as shown in Figure~\ref{Fig_evol_12390}. The time evolution of helicity injection $dH/dt$ is positive definite quantity till July 28 which then turns to negative values. EUV flux of the AR in 94~\AA~and GOES X-ray flux are plotted in Figure~\ref{Fig_evol_12390}c, which indicate that the AR exhibits very low flaring activity limited to B-class flares. 

\begin{figure*}
	\centering
	\includegraphics[width=.99\textwidth,clip=]{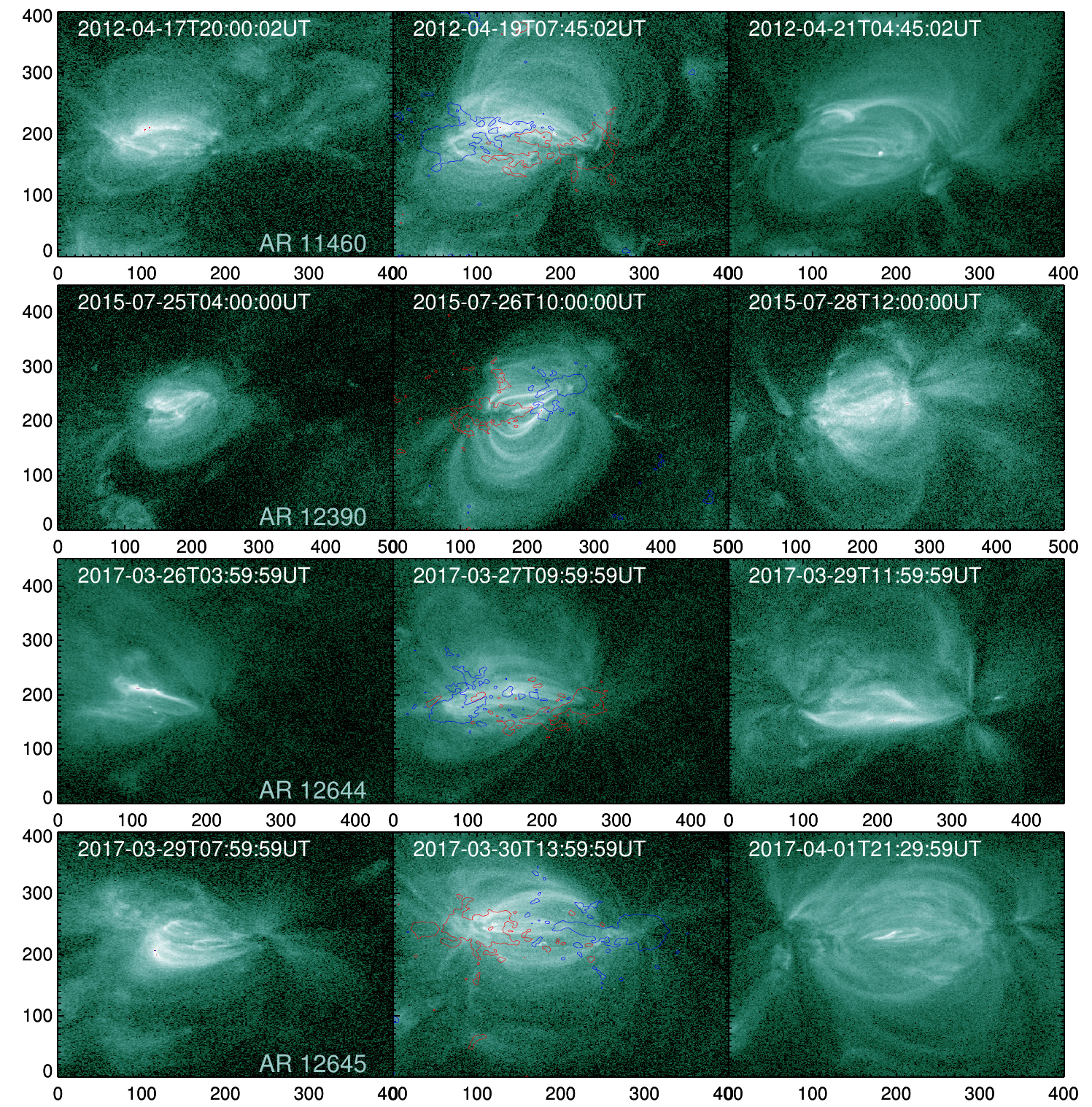}
	\caption{Observations of the AR corona in AIA 94~\AA~. Images at different times (left to tight) are displayed for each AR. Contours of $B_z$ at $\pm120$ G (blue/red) are shown in the middle panels. The images in each AR exhibit plasma loops connecting the simple bipolar flux distribution (leading and following polarities) like a mildly sheared and/or potential arcade magnetic structure. The axes units in each image are in pixels of 0''.6 size. To support this figure, a movie is available in the Electronic Supplementary Materials. The movie shows the full evolution of the ARs in the images of AIA 94~\AA~ channel.}
	\label{Fig_aia}
\end{figure*}

\subsection{AR 12644}
The AR 12644 started emerging on March 25, 2017 at the disc location E50$^o$N15$^o$. Following the Hale polarity law, a leading negative sunspot was seen to grow in the course of AR emergence and further evolution. We analyzed the observations of this AR during March 25-30. In Figure~\ref{Fig_evol_12644}, the time evolution of the magnetic and emission parameters are plotted. The magnetic flux of the both polarities increase till March 28 which then show a slight decrease behavior. Similar to earlier two ARs, the time evolution of the $dH/dt$ in this AR exhibits sign-change nature from positive to negative values starting from March 28. This trend of $dH/dt$ is also consistent with the average twist parameter $\alpha_{av}$. The GOES X-ray flux and EUV 94~AA~flux profiles are shown in Figure~\ref{Fig_evol_12644}c, where some of the peaks in EUV 94 flux are co-temporal with the GOES X-ray flux. We noticed that the overall flaring activity from this AR is limited to C-flares without CMEs.

\subsection{AR 12645}
The AR 12645 started emerging on March 27, 2017 at the disc location E60$^o$S10$^o$. The leading polarity is positive as for the most ARs in the southern hemisphere. In order to minimize projection effects in our calculations of the magnetic parameters, we analyzed the observations of this AR during March 27 to April 3. As shown in Figure 5a, the flux content keeps growing as the AR emerges and spreads in spatial extent of flux distribution. Unlike the previous AR, the $dH/dt$ evolution turns from negative sign values to positive from mid of March 31, which is also consistent with the average twist parameter $\alpha_{av}$. As shown in Figure~\ref{Fig_evol_12645}c, few of the peaks in EUV flux are co-temporal with the peaks in GOES X-ray flux, which are identified as C-flares without being associated with CMEs. 

\section{Discussion and Conclusion}
\label{Dis}

We present the observations of four ARs with photospheric flux motions generating successive injections of opposite signs of MH. Such ARs are outliers disobeying the hemispheric helicity rule \citep{rust1996,pevtsov1995,Pevtsov2008, Liu2014_HemisphericRule} when evaluated from the observations of the on-disk ARs. These ARs are somewhat not common, especially of interest to shed light on the role of conservation property of magnetic helicity for the launch of CMEs. From the point of helicity accumulation, the ARs with change of $dH/dt$ sign over time may not favor the twisted flux rope formation implying to a state of smaller content of coronal helicity below the eruptive threshold, therefore a CME eruption is unlikely \citep{zhangmei2005,Zhangmei2006_MagFld_confin}. On this line, recent theoretical and numerical studies also propose the existence of a helicity upper bound for the launch of a CME from an AR \citep{Zhangmei2008_HelBound,Zuccarello2018_Thres_MagHel}.

ARs of predominant sign of $dH/dt$ or strong $\alpha_{av}$ indicate strong twisted magnetic structures and typically appear as sigmoidal shape in X-ray or EUV observations \citep{canfield1999,Green2007, RomanoP2014, Vemareddy2016_sunspot_rot,  Vemareddy2017_homologus, vemareddy2019_VeryFast, DhakalS2020}. Since the $dH/dt$ is not with a predominant sign over the studied period of observation, the ARs in this study have not built up twisted structures (sigmoidal shapes). To this end, we analyzed the time-sequence images in AIA 94~\AA~channel. In Figure~\ref{Fig_aia}, representative images of AIA 94 \AA~observations of the AR corona at three different times are displayed. As noticed, the coronal images exhibit plasma loops connecting the simple bipolar flux distribution (leading and following polarities) like a mildly sheared and/or potential arcade magnetic structure. These loops have no signatures of sigmoidal shape (Figure~\ref{Fig_aia} and supported animation), as observed in many erupting ARs \citep{canfield1999, gibson2002, Green2002, Vemareddy2017_homologus}. Inspection of motion images imply intermittent brightenings some of which are identified as C-flares in GOES X-ray flux classification. In addition, we have also examined the LASCO white-light images\footnote{\url{https://cdaw.gsfc.nasa.gov/CME_list/UNIVERSAL/}} in conjunction with the AR evolution in AIA 94 images, which also delineate the absence of associated CMEs with the studied ARs. 

In terms of eruptive nature, the magnetic flux in the studied ARs is small compared to super ARs (e.g., \citealt{ChenAnqin2016_SARs}). Since the helicity flux scales with the square of the net flux ($\Phi^2$) in the AR, normalizing the helicity flux with net average flux ($dH/dt/\Phi^2$)is useful as suggested by \citet{Vemareddy2017_homologus,vemareddy2019_VeryFast}. This quantity indicates the overall twisted-ness of the AR, where a smaller value of normalized helicity refers to a large fraction of the flux being untwisted and acting as background confined field. Such an idea of normalizing non-potential parameters by magnetic flux was used in the recent statistical studies of \citet{TingLi2020_MagneticFlux,TingLi2022_MagneticParameter} to find the CME association with the flares. They suggested that the flares of the same GOES class from a larger flux ARs are unlikely to be eruptive. However, ARs of sign changing helicity are very fundamental in that they are unlikely to form flux ropes in the first place. Therefore, in order to disentangle the effect of background field from the large value of helicity flux generated from flux motions, it is important to normalize the helicity flux from ARs of small to large magnetic flux content. Also, we need to find observational evidences of sign-changing helicity flux in several confined ARs of different size.

Further, the studied ARs have not developed long sheared polarity inversion (SPIL) in the course of their evolution. Several studies showed that the erupting ARs contain strong SPILs over which twisted FR forms. These erupting ARs have non-neutralized currents spread along the PIL such that the net current is not balanced in a given polarity \citep{Torok2014,Vemareddy2019_DegEle,Kazachenko2022_MagFieldFlareRib}. In our ARs, soon after the emergence, the opposite polarities separate each other so that SPIL cannot form. This study provides the characteristics of helicity flux evolution in ARs that are incapable of producing eruptions, and more such studies would help to quantify the eruptive capability of a given AR.

\section*{Acknowledgements}
The data have been used here courtesy of NASA/SDO and HMI science team. We thank the HMI science team for the open data policy of processed vector magnetograms. I thank the anonymous referee for helpful comments and suggestions.

\section*{Data Availability} The data used in this manuscript is from NASA's SDO mission and is publicly available from Joint Science Operations Center (
\url{http://jsoc.stanford.edu/}).

\bibliographystyle{mnras.bst} 

%
%

\bsp	
\label{lastpage}
\end{document}